\documentclass{scrartcl}
\usepackage[english]{babel}
\usepackage[latin1]{inputenc}
\usepackage{amsmath, amsthm, amssymb}
\usepackage[pdftex]{graphicx}
\usepackage{amssymb}
\usepackage{latexsym}
\usepackage{amstext}
\newcommand{\ad}{\mathbf{a_2}}
\newcommand{\adad}{\mathbf{a_2 \oplus a_2}}
\newcommand{\adadud}{\mathbf{a_2^{(1)} \oplus a_2^{(2)}}}
\newcommand{\adc}{\mathbf{a_2^c}}
\newcommand{\add}{\mathbf{a_2^{(2)}}}
\newcommand{\adf}{\mathbf{a_2^f}}
\newcommand{\adgd}{\mathbf{a_2^{g_2}}}
\newcommand{\adgu}{\mathbf{a_2^{g_1}}}
\newcommand{\adu}{\mathbf{a_2^{(1)}}}

\newcommand{\cc}{\mathbf{C}}
\newcommand{\ekm}{\varepsilon_k^-}

\newcommand{\ekp}{\varepsilon_k^+}
\newcommand{\ekpm}{\varepsilon_k^{\pm}}

\newcommand{\eo}{\mathbf{e_8}}
\newcommand{\es}{\mathbf{e_6}}
\newcommand{\est}{\mathbf{e_7}}
\newcommand{\fq}{\mathbf{f_4}}
\newcommand{\gd}{\mathbf{g_2}}
\newcommand{\gon}{\mathbf{g_0^n}}
\newcommand{\hu}{\mathbf{H}}
\newcommand{\jo}{\mathbf{J}}
\newcommand{\job}{\mathbf{\bar J}}
\newcommand{\lk}{\mathfrak{L}}
\newcommand{\oo}{\textbf{\large $\mathfrak C$}}
\newcommand{\qq}{\mathbf{Q}}
\newcommand{\rr}{\mathbf{R}}
\newcommand{\str}{\text{str}}
\newcommand{\sut}{\mathbf{su(3)}}

\numberwithin{equation}{section}
\begin{document}
%
\begin{center}
\Large
\textbf{Exceptional Lie Algebras, SU(3) and Jordan Pairs}

\vspace{1cm}
\large\textbf{Piero Truini}\\
\normalsize\textit{Dipartimento di Fisica, Universit\` a degli Studi\\ via Dodecaneso 33, I-16146 Genova,  Italy}
\end{center}

\vspace{.5cm}
\abstract{
\textbf{Abstract.}\textit{A simple unifying view of the exceptional Lie algebras is presented. The underlying Jordan pair content and role are exhibited. Each algebra contains three Jordan pairs sharing the same Lie algebra of automorphisms and the same external su(3) symmetry. Eventual physical applications and implications of the theory are outlined. }}

\vspace{1cm}
\normalsize
\section{Introduction}
The main purpose of this paper is to exhibit a unifying view of all exceptional Lie algebras, which is also very intuitive from the point of view of  elementary particle physics. The result is represented by the following diagram:

\begin{center}
\includegraphics{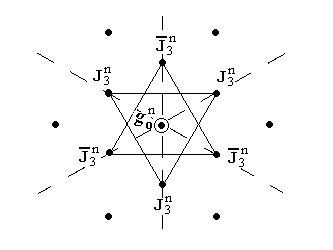}
\end{center}
\begin{center}\textbf{Fig.1} A unifying view of the exceptional Lie algebra roots\end{center}

It is a very simple, highly intuitive unifying view of all Exceptional Lie algebras and we will use it repeatedly to unfold the largest algebra $\mathbf{e_8}$. The picture shows the Jordan pair content of each exceptional Lie algebra: there are three Jordan pairs $(\mathbf{J,\bar J})$, each of which lies on an axis symmetrically with respect to the center of the diagram. Each pair doubles a Jordan algebra with involution (the conjugate representation). The three Jordan algebras (and their conjugate) globally behave like a $\mathbf 3$ (and a $\mathbf{\bar 3}$) dimensional representation of the outer $\sut$. The algebra denoted by $\mathbf{g_0^n}$ in the center (plus the Cartan generator associated with the axis along which the pair lies) is the algebra of the automorphism group of the Jordan Pair (the structure group of the corresponding Jordan Algebra). In the case of  $\eo$, the algebra $\mathbf{g_0^n}$ is $\es$,  described by a similar diagram, and we can iterate the process. What we eventually end up with is a decomposition of $\eo$ entirely given in terms of Jordan pairs and $\mathbf{su(3)}$'s. 

The interest of physicists on the exceptional Lie algebras, and on $\eo$ in particular, has a long standing tradition, starting from the pioneering work of G\"ursey \cite{gursey80} on Grand Unification, \cite{gs}-\nocite{crem}\nocite{bt1}\nocite{chsw}\nocite{gross}\nocite{ferrara1}\cite{ferrara2}. In the effort of unifying all interactions in a consistent quantum theory that includes gravity, the most successful model of string theory is based on  $\mathbf{e_8}$ and an alternative theory known as \emph{loop quantum gravity} (see \cite{carlo} for an excellent and comprehensive review) has also led towards the Exceptional algebras, and $\mathbf{e_8}$ in particular \cite{smoli}.

There is a wide consensus in both mathematics and physics on the appeal of the largest exceptional Lie algebra $\eo$, considered by many beautiful in spite of its complexity. A best synthesis of this is in B. Kostant's words: \emph{It is easy to arrive at the feeling that a final understanding of the universe must somehow involve E(8), or otherwise put, nature would be foolish not to utilize E(8).} 

Kostant defines $\eo$ \emph{a symphony of 2, 3, 5}. In the more modest view of the exceptional algebras I present here the numbers 1, 2, 3 play the central role: they govern the structure. Number 1 is the whole, the universe of the theory: a Lie algebra. Number 2 stands for pair, and we view it as particle-antiparticle duality represented by Jordan pairs. Number 3 is the number of colors and the number of families: each Jordan pair occurs three times, in a $\sut$ symmetry. That is all you need in order to build $\mathbf{e_8}$, as we are going to show.
 
\section{Jordan Pairs} In this section we review the concept of a Jordan Pair, \cite{loos1} (see also \cite{mccrimm1} for an enlightening overview). Jordan Algebras have traveled a long journey, since their appearance in the 30's \cite{jvw}. The modern formulation \cite{jacob1} involves a quadratic map $U_x y$ (like $xyx$ for associative algebras) instead of the original symmetric product $x \circ y = xy + yx$. The quadratic map and its linearization $V_{x,y} z = (U_{x+z} - U_x - U_z)y$ (like $xyz+zyx$ in the associative case) reveal  the mathematical structure of Jordan Algebras much more clearly, through the notion of inverse, inner ideal, generic norm, etc. The axioms are:
\begin{equation}
U_1 = Id \quad , \qquad
U_x V_{y,x} = V_{x,y} U_x \quad  , \qquad
U_{U_x y} = U_x U_y U_x
\label{qja}
\end{equation}
The quadratic formulation led to the concept of Jordan Triple systems \cite{myb}, an example of which is a pair of modules represented by rectangular matrices. There is no way of multiplying two matrices $x$ and $y$ , say $n\times m$ and $m\times n$ respectively, by means of a bilinear product. But one can do it using a product like $xyx$, quadratic in $x$ and linear in $y$. Notice that, like in the case of rectangular matrices, there needs not be a unity in these structures. The axioms are in this case:
\begin{equation}
U_x V_{y,x} = V_{x,y} U_x \quad  , \qquad
V_{U_x y , y} = V_{x , U_y x} \quad , \qquad
U_{U_x y} = U_x U_y U_x
\label{jts}
\end{equation}

Finally, a Jordan Pair, is just a pair of modules $(V^+, V^-)$ acting on each other (but not on themselves) like a Jordan Triple:
\begin{equation}\begin{array}{ll}
U_{x^\sigma} V_{y^{-\sigma},x^\sigma} &= V_{x^\sigma,y^{-\sigma}} U_{x^\sigma} 
\\
V_{U_{x^\sigma} y^{-\sigma} , y^{-\sigma}} &= V_{x ^\sigma, U_{y^{-\sigma}} x^\sigma} \\
U_{U_{x^\sigma} y^{-\sigma}} &= U_{x^\sigma} U_{y^{-\sigma}} U_{x^\sigma}\end{array}
\label{jp}
\end{equation}
where $\sigma = \pm$ and $x^\sigma \in V^{+\sigma} \, , \; y^{-\sigma} \in V^{-\sigma}$.

Jordan pairs are strongly related to the Tits-Kantor-Koecher construction of Lie Algebras \cite{tits1}-\nocite{kantor1}\cite{koecher1} (see also the interesting relation to Hopf algebras, \cite{faulk}):
\begin{equation}
\lk = J \oplus \str(J) \oplus \bar{J} \label{tkk}
\end{equation}
where $J$ is a Jordan algebra and $\str(J)= L(J) \oplus Der(J)$ is the structure algebra of $J$ \cite{mccrimm1}; $L(x)$ is the left multiplication in $J$: $L(x) y = x \circ y$ and $Der(J) = [L(J), L(J)]$ is the algebra of derivations of $J$ (the algebra of the automorphism group of $J$) \cite{schafer1}\cite{schafer2}.

 In the case of (complex) exceptional Lie algebras this construction applies to $\est$, with $J = \jo^8_3$, the 27-dimensional exceptional Jordan algebra of $3 \times 3$ hermitean matrices over the octonions, and $\str(J) = \es \otimes \cc$ ($\mathbf{C}$, the complex field). The algebra $\es$ is called the \emph{reduced structure algebra} of $J$ , $\str_0(J)$, namely the structure algebra with the generator corresponding to the multiplication by a complex number taken away: $\es = L(J_0) \oplus Der(J)$, with $J_0$ denoting the traceless elements of $J$.

The Tits-Kantor-Koecher construction can be generalized as follows: if $\lk$ is a three graded Lie algebra:
\begin{equation} \lk = \lk_{-1} \oplus L_0 \oplus \lk_1 \qquad [\lk_i , \lk_j] \subset \lk_{i+j} \end{equation}
so that $ [\lk_i , \lk_j] = 0$ whenever $|i+j|>1$, then $(\lk_1, \lk_{-1})$ forms a Jordan pair, with the Jacobi identity forcing the elements of the pair to act on each other like in Jordan triple system. The link with the Tits-Kantor-Koecher construction is obtained by letting $J = \lk_1$, $\bar J = \lk_{-1}$. The structure group of $J$ is the automorphism group of the Jordan pair $(J, \bar J)$ and the trilinear product $ V_{x^\sigma , y^{-\sigma}} z^\sigma$ is
\[ V_{x^\sigma , y^{-\sigma}} z^\sigma = [[x^\sigma , y^{-\sigma}], z^\sigma]\]

\section{The Freudenthal-Tits Magic Square}

\begin{table}
\begin{center}
\begin{tabular}{c|cccc}
$\hu \setminus \jo$  & $J_3^1$ & $J_3^2$ & $J_3^4$ & $J_3^8$\\
\hline
$\rr$ & $\mathbf{a_1}$ & $\mathbf{a_2}$ & $\mathbf{c_3}$  & $\fq$ \\
$\cc$ & $\mathbf{a_2}$  & $\adad$  & $\mathbf{a_5}$  & $\es$ \\
$\qq$ & $\mathbf{c_3}$  & $\mathbf{a_5}$  & $\mathbf{d_6}$  & $\est$ \\
$\oo$ & $\fq$  & $\es$  & $\est$  & $\eo$\\
\end{tabular}
\vspace{.4cm}\\
Table 1. The Freudenthal-Tits Magic Square
\end{center}
\end{table}
The theory of exceptional Lie algebras has had a major advance with the development of two related objects: the Tits construction and the Freudenthal-Tits Magic Square \cite{tits2, freu1}. The Freudenthal-Tits Magic Square is a table of Lie algebras related to both Jordan algebras and Hurwitz algebras $\mathbf{H}$, namely the algebras of real ($\rr$), complex ($\cc$), quaternion ($\qq$) and octonion or Cayley ($\oo$) numbers. In particular the Jordan Algebras involved in the Magic Square are the algebras of $3\times 3$ hermitean matrices over $\hu$.
\begin{equation}
\left( \begin{array}{ccc} \alpha & a & \bar b \\ \bar a & \beta & c \\ b & \bar c & \gamma \end{array} \right) \qquad \alpha \, , \beta \, , \gamma \in \cc \; ; \, a\, , b \, , c \in \hu 
\label{jm}
\end{equation}
 We denote them by $J_3^n$ where $n=1,2,3,4$ for $\hu = \rr , \, \cc  ,\, \qq  ,\, \oo $ respectively. In this paper only complex Lie algebras are considered. Therefore each algebra $\hu$ is over the complex field and the $a \to \bar a$ conjugation in (\ref{jm}) changes the signs of the imaginary units of $\hu$ but does not conjugate the imaginary unit of the underlying complex field. The Freudenthal-Tits Magic Square is shown in Table 1.

The way the Magic Square is constructed is due to Tits:
\begin{equation}
\lk = Der(\mathbf{H}) \oplus (\mathbf{H}_0 \otimes \mathbf{J_0}) \oplus Der(\mathbf{J}) \label{titsms}
\end{equation}

Here the subscript $0$ stands for traceless. $Der(\hu)$ is the algebra of derivations of $\hu$, which is nothing for $\hu = \rr \, , \cc$, whereas $Der(\qq) = \mathbf{a_1}$ and $Der(\oo) = \gd$.

We also have the following tight link between the entries of the Magic Square and Jordan structures.

The Lie algebras $\mathbf{g_I}$ in the first row of the Magic Square are the algebras of derivations of the Jordan Algebra in the same column (the corresponding group is the automorphism group the Jordan algebra):
\[\mathbf{g_I} = Der(\jo)\quad \text{namely :}\]  
\[ \mathbf{a_1} = Der(\jo_3^1 )\; , \;  \mathbf{a_2} = Der(\jo_3^2 )\; , \;  \mathbf{c_3} = Der(\jo_3^4) \; , \; \fq = Der(\jo_3^8 )\; , \; \]

The Lie algebras $\mathbf{g_{II}}$  in the second row are the reduced structure algebras of the Jordan Algebra in the same column
\[\mathbf{g_{II}} = \str_0(\jo)\quad \text{namely :}\]
\[ \mathbf{a_2} = \str_0(\jo_3^1) \; , \;  \adad = \str_0(\jo_3^2) \; , \;  \mathbf{a_5} = \str_0(\jo_3^4) \; , \; \es = \str_0(\jo_3^8) \; , \; \]

The Lie algebras $\mathbf{g_{III}}$ in the third row are three graded and can be written via the Tits-Kantor-Koecher construction \eqref{tkk} or in terms of generalized $2 \times 2$ matrices \cite{tob}:
\[\mathbf{g_{III}} = \jo \oplus (\mathbf{g_{II}}\otimes \cc ) \oplus \job\quad \text{namely :}\]
\begin{align*}
\mathbf{c_3} = \jo_3^1 \oplus (\ad \oplus \cc) \oplus \bar \jo_3^1\; &, \; \mathbf{a_5} = \jo_3^2 \oplus (\adad \oplus \cc) \oplus \bar \jo_3^2\\
\mathbf{d_6} = \jo_3^4 \oplus (\mathbf{a_5} \oplus \cc) \oplus \bar \jo_3^4\; &, \; \est = \jo_3^8 \oplus (\es \oplus \cc) \oplus \bar \jo_3^8
\end{align*}

In our opinion the most natural way of extending a similar analysis to the fourth row is the one described in this paper and represented in Fig. 1 or in the expression \ref{uv} of the next section.

\section{The Jordan Pairs inside the Exceptional Lie Algebras}

In this section we work with the roots of the exceptional Lie algebras and postpone the discussion on explicit representations of the generators to the next section. 

We first fix the notation and show the explicit set of roots we will use, \cite{bour}.

This is done in Table 2, where $\{ k_i , \, i = 1, \dots ,8 \}$ denotes an orthonormal basis in $\rr ^8$. 

\begin{table}
\begin{equation*}
\begin{array}{|l|l|c|}
\hline
\lk &\text{roots}\;\{ k_i , \, i = 1, \dots 8 \}$ an orthonormal basis in $\rr ^8 & \text{n. of roots}\\
\hline
\gd & &\textbf{12 roots} \\
&(k_i - k_j) \qquad  i\ne j = 1, 2, 3 &  6 \\
&\pm \frac{1}{3} (- 2 k_i + k_j + k_l) \qquad  i\ne j  \ne l = 1, 2, 3 & 6\\
\hline
\fq & & \textbf{48 roots}\\
& \pm k_i  \qquad  i = 1,\dots , 4 & 8\\
&\pm k_i \pm  k_j \qquad  i\ne j = 1, \dots ,4 & 4\times \binom{4}{2} = 24 \\
&\frac{1}{2} (\pm k_1 \pm k_2 \pm k_3 \pm k_4) & 2^4 = 16\\
\hline
\es & &\textbf{72 roots}\\
&\pm k_i \pm  k_j \qquad  i\ne j = 1, \dots ,5 &  4\times \binom{5}{2} = 40 \\
&\frac{1}{2} (\pm k_1 \pm k_2 \pm k_3 \pm k_4  \pm k_5  \pm \sqrt{3}  k_6)^*  & 2^5= 32\\
& ^*\text{ [odd number of + signs]} & \\
\hline
\est & &\textbf{126 roots}\\
& \pm \sqrt{2} k_7&  2\\
&\pm k_i \pm  k_j \qquad  i\ne j = 1, \dots ,6 & 4\times \binom{6}{2} = 60 \\
&\frac{1}{2} (\pm k_1 \pm k_2 \pm k_3 \pm k_4  \pm k_5  \pm k_6 \pm \sqrt{2}  k_7)^*   &2^6= 64\\
&^*\text{[even number of + signs]} & \\
\hline
\eo & &\textbf{240 roots}\\
&\pm k_i \pm  k_j \qquad i\ne j = 1, \dots ,8 &4\times \binom{8}{2} = 112 \\
&\frac{1}{2} (\pm k_1 \pm k_2 \pm k_3 \pm k_4  \pm k_5  \pm k_6 \pm k_7 \pm k_8)^*  & 2^7= 128\\
&^*\text{[even number of + signs]} & \\
\hline
\end{array}
\end{equation*}
\begin{center} Table 2. The roots of the Exceptional Lie Algebras\end{center}
\end{table}

We then place the roots, case by case, on a figure like Fig. 1, which is nothing but the projection of all roots on the plane of an $\ad$ sub-algebra  (we use the standard notation $\ad$ for the complexification of $\sut$) .
 Notice that $\gd$ itself has a root diagram represented by the same dots appearing in Fig.1. 

We start therefore with it and place the roots shown in Table 2 on the external hexagon of $\ad$ and on its $(3, \bar 3)$. 

The result is shown in Fig.2. 

\begin{figure}
\begin{center}
\includegraphics{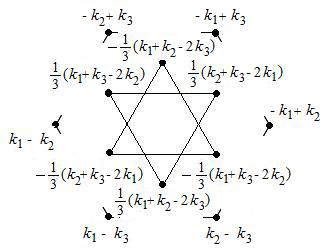}
\end{center}
\begin{center}\textbf{Fig.2} Roots of $\gd$\end{center}
\end{figure}

We project on the same plane $\mathbf\Pi$ spanned by the $\ad$ roots $(k_i - k_j)\, ,\; i,j = 1,2,3$, the roots of all exceptional Lie algebras and find that the root diagrams of $\fq$, $\es$ ,$\est$ and $\eo$,  are as in Fig. 3.
The notation for the Jordan algebras in the figure is the same used in Table 1 for the Freudenthal-Tits Magic Square: $\mathbf{J_3^n}\, , n=1,2,4,8$ is the Jordan algebra of $3\times 3$ hermitean matrices over $\rr$, $\cc$, $\qq$, and $\oo$  respectively. The algebra $\mathbf{g_0^n} \, , n = 1,2,4,8$ is $\ad$, $\adad$, $\mathbf{a_5}$ and $\es$ respectively; $\mathbf{g_0^n} \oplus \cc$ is the algebra of the automorphism group of each Jordan pair $\mathbf{V^n} = (\jo_3^n , \job _3^n)$. We associate roots to Jordan pairs and we check that the projection of these roots lie along an axis, symmetrically with respect to the center. The $\cc$ in  $\mathbf{g_0^n} \oplus \cc$ stands for the complex field and represents the action on $\mathbf{V^n}$ (multiplication by a complex number) of the Cartan generator associated with that axis.

\begin{figure}
\begin{center}
\includegraphics{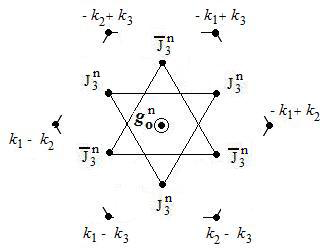}\\ \nopagebreak
\textbf{Fig.3} Roots of $\fq$, $\es$ ,$\est$ and $\eo$,  for $n = 1,2,4,8$ respectively, \\  projected on the plane $\mathbf\Pi$
\end{center}
\end{figure}

If we write $\lk^n$ for  $\fq$, $\es$ ,$\est$ and $\eo$, $n=1,2,4,8$,  we get the unifying expression 

\begin{equation}\lk^n = \ad \oplus \gon \oplus 3 \times(\jo_3^n, \bar \jo _3^n)\qquad \text{where}\quad \gon = \str_0(\jo_3^n)\label{uv}\end{equation}

This is not only a unifying view of the exceptional Lie algebras, but also, in our opinion, a natural way of looking at the fourth row of the Magic Square. Notice that $\gon$ is the Lie algebra in the second row ($\mathbf{g_{II}}$), at  the same column of $\lk^n$ and that $\gon \oplus \cc \oplus \mathbf{V^n}$ is the Lie algebra in the third row ($\mathbf{g_{III}}$),  same column, for any of the three Jordan pairs $\mathbf{V^n}$ in $\lk^n$ (Fig. 4).

\begin{figure}
\begin{center}
\includegraphics{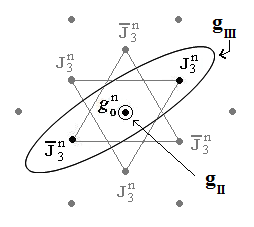}\\ \nopagebreak
\textbf{Fig.4} Roots of the Exceptional Lie algebras with $\mathbf{g_{II}}$ and $\mathbf{g_{III}}$  highlighted.
\end{center}
\end{figure}

We will explicitly show the roots associated with a Jordan algebra in Fig.3. In particular we will pick the one whose projection on the plane $\mathbf \Pi$ is $\frac{1}{3}(k_2 + k_3 - 2 k_1)$ (see Fig. 2 for this vector), that is the highest weight in the 3-dimensional representation of $\sut \sim \ad$. We will refer to this Jordan algebra as the highest weight (h.w.) $\jo_3^n$. 
The other Jordan algebras are obtained by a permutation of indexes and their conjugate ones by a change of sign.

Let us explain why we say that certain roots correspond to a Jordan pair. The reason lies in the Tits-Kantor-Koecher construction \eqref{tkk}, which is related to the third row of the Freudenthal-Tits Magic Square. There is only one way of realizing the embedding a $\mathbf{g_{II}}\subset \mathbf{g_{III}} \subset \lk^n$ so that the $(\jo_3^n,\job_3^n)$ modules for $\mathbf{g_{II}}$ lie on parallel spaces at the same distance along a fixed axis. This is precisely the way we will describe the Jordan pair content of the algebras and this shows the uniqueness of the construction. We know from the three grading structure of  $\mathbf{g_{III}}$ that the pair $(\jo_3^n , \job_3^n)$ is indeed a Jordan Pair and that $\str(\jo_3^n)=\mathbf{g_{II}} \oplus \cc$  is the Lie  algebra of the automorphism group of the Jordan pair. This proves that the Jordan structures we have referred to so far are indeed so.

We now examine the four exceptional algebras $\fq$, $\es$ ,$\est$ and $\eo$ case by case. For each we will show:

\begin{enumerate}
\item the roots associated with the h.w. Jordan algebra $\jo_3^n$
\item the roots associated with $\gon =\mathbf{g_{II}} $
\item the geometry of the Jordan pair $\mathbf V$ and of the algebra  $\mathbf{g_{III}}=\gon \oplus \cc \oplus \mathbf V$
\item nested Jordan pairs
\end{enumerate}

\subsection{\Large $\;\;\fq$}

\subsubsection{The roots associated with the h.w. Jordan algebra $\jo_3^1$}

The 6 roots corresponding to the highest weight (h.w.) $\jo_3^1$ in Fig. 3 are

\begin{equation}
\begin{array}{l}
- k_1 \qquad - k_1 \pm k_4 \\
\frac{1}{2} ( - k_1 + k_2 + k_3 \pm k_4) \\
k_2 + k_3
\end{array}
\end{equation}

All these vectors have projection $\frac{1}{3}(k_2 + k_3 - 2 k_1)$ on the plane $\mathbf{\Pi}$, namely the plane perpendicular to the vector $ (k_1 + k_2 + k_3)$  in the 3-dim space spanned by  $k_1$, $k_2$, $k_3$.

The other 2 $\jo_3^1$'s are obtained by a cyclic permutation of the indexes of  $k_1$, $k_2$, $k_3$. The 3  $\job_3^1$'s are obtained by reversing the roots' sign.

\subsubsection{The roots associated with $\mathbf{g_{II}} =  \mathbf{g_0^1}$}

The 6 roots of   $\mathbf{g_0^1} = \ad$ are 

\begin{equation}
\pm k_4 \qquad \pm \frac{1}{2} ( k_1 + k_2 + k_3 \pm k_4) 
\end{equation}

Notice the scaling by $\frac{\sqrt{2}}{2}$ of these roots' norm with respect to the roots' norm of  the $\ad$ in Fig.3.

\subsubsection{The geometry of the Jordan pair $\mathbf V$ and of  $\mathbf{g_{III}}=\mathbf{g_0^1} \oplus \cc \oplus \mathbf V$}

The root vectors of the h.w. $\jo_3^1$  all lie on the plane defined by the linear span:
\[
\mathbf\Pi^+ = \left\{ \frac{1}{3} (k_2 + k_3 - 2 k_1) + s\ k_4 + t\ \frac{\sqrt{3}}{3} (k_1 + k_2 + k_3) \right\}
\]
with $t= -  \frac{\sqrt{3}}{3} \; , s= 0, \pm 1 \; ; \; t= \frac{\sqrt{3}}{6} \;  , s= \pm \frac{1}{2} \; ; \; t= \frac{2 \sqrt{3}}{3} \; , s= 0$ . 

The plane $\mathbf\Pi^+ $  is parallel to  plane $\mathbf\Pi^0$:
\[
\mathbf\Pi^0 = \left\{ s\  k_4 + t\ \frac{\sqrt{3}}{3} (k_1 + k_2 + k_3) \right\}
\]
on which the $\mathbf{g_0^1}$ roots lie, for $t= 0 \; , s= \pm 1$ and for $t= -  \frac{\sqrt{3}}{2} \;  , s= \pm \frac{1}{2}$
and to the plane
\[
\mathbf\Pi^- = \left\{ - \frac{1}{3} (k_2 + k_3 - 2 k_1) + s\ k_4 + t\ \frac{\sqrt{3}}{3} (k_1 + k_2 + k_3) \right\}
\]
on which the roots of  the $\job_3^1$ opposite to the h.w. $\jo_3^1$ lie for $t= \frac{\sqrt{3}}{3} \; , s= 0, \pm 1 \; ; \; t= -  \frac{\sqrt{3}}{6} \;  , s= \pm \frac{1}{2} \; ; \; t= - \frac{2 \sqrt{3}}{3} \; , s= 0$ . 

Both planes $\mathbf{\Pi^\pm}$ have the same distance $\frac{\sqrt{6}}{3}$  from $\mathbf{\Pi^0}$ .  These two $\jo_3^1$ form a Jordan pair of conjugate $\ad$-representations $(6, \bar 6)$.

The roots on the three planes form the root diagram of $\mathbf{c_3}$ as shown in Fig.5.

\begin{figure}
\begin{center}
\includegraphics{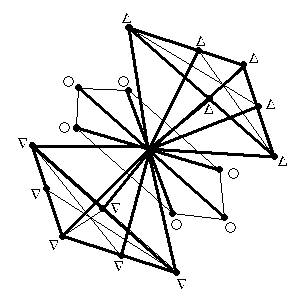}\\
\textbf{Fig.5} Root digram of $\mathbf{c_3}$ showing $\ad$ and the Jordan pair $(6,\bar 6)$\end{center}
\end{figure}

This Jordan pair is clearly visible in the figure. The Lie Algebra of the automorphism group of this pair is $\ad \oplus \cc$ where $\cc$ is the complex linear span of the Cartan generator associated with the axis along the vector $\frac{1}{3} (k_2 + k_3 - 2 k_1)$ which is precisely the direction of the Jordan pair in Fig 3. All the points of  the h.w. $\jo_3^1$ (resp.$\job_3^1$, opposite to it with respect to the center of Fig.3) project on the point $\frac{1}{3} (k_2 + k_3 - 2 k_1)$ (resp. $- \frac{1}{3} (k_2 + k_3 - 2 k_1)$) in the plane of Fig. 3.

There is only one way of embedding a $\mathbf{c_3}$ sub-algebra within $\fq$ so that the $(6,\bar 6)$ modules for $\ad$ lie on parallel planes at the same distance along a fixed axis. This is precisely the way we have described above and this shows its uniqueness. We know from the three grading structure of  $\mathbf{c_3}$ that the pair $(\jo_3^1 , \job_3^1)$ is indeed a Jordan Pair and that $\str(\jo)$ ($=\ad \oplus \cc$ in this case) is the Lie  algebra of the automorphism group of the Jordan pair. 

Still, one would like to see explicit representations of the generators related to roots, on one side, and Jordan algebra or pair elements on the other side. We will approach this topic in the next section. 

Similar proofs of the fact that we are dealing with Jordan structures associated with roots applies for the other exceptional Lie algebras that we are going to examine next.

By a cyclic permutation of  the indexes of  $k_1$, $k_2$, $k_3$ we obtain an analogous result for the other two Jordan pairs, all sharing the same $\ad$ roots for the algebra $\mathbf{g_0^1}$, but with different orientations of the axis defining $\cc$ along the vectors  $\frac{1}{3} (k_1 + k_3 - 2 k_2)$  and  $\frac{1}{3} (k_1 + k_2 - 2 k_3)$ . We get in four dimensions three copies of $\mathbf{c_3}$ all sharing the same $\ad$. All the planes spanned by the three Jordan pairs are parallel to the plane $\mathbf\Pi^0$, and all at the same distance from it. 

Notice that in four dimensions there is an infinite number of planes parallel to a given one, all at the same distance from it. In fact, given an orthonormal basis $\{ \mathbf b_1 , \mathbf b_2 , \mathbf b_3 , \mathbf b_4\}$ in $\rr^4$, a plane $\mathbf P = \{ \alpha \mathbf b_1 + \beta \mathbf b_2 \}$, and a distance $d$, the infinite planes parallel to $\mathbf P$ at a distance $d$ are $\{ \mathbf v + s \mathbf b_1 + t \mathbf b_2 \}$, where $\mathbf v = x \mathbf b_3 + y \mathbf b_4$ and $x^2 + y^2 = d^2$.

\subsubsection{Nested Jordan pairs}
If we dig inside $\mathbf{g_0^1}$ we find another Jordan Pair plus the Lie algebra of its automorphism group: these are a  $(2, \bar 2)$ of $\mathbf{a_1}$ plus $\mathbf{a_1} \oplus \cc$ making up, all together,  $\ad$.

\subsection{\Large $\;\;\es$}

\subsubsection{The roots associated with the h.w. Jordan algebra $\jo_3^2$}

The 9 roots corresponding to the highest weight (h.w.) $\jo_3^2$ in Fig. 3 are

\begin{equation}
\begin{array}{l}
- k_1 \pm k_4  \; , \qquad - k_1 \pm k_5 \; , \qquad k_2 + k_3\\
\frac{1}{2} ( - k_1 + k_2 + k_3 + k_4 - k_5 - \sqrt{3}  k_6)  \; ,\quad \frac{1}{2} ( - k_1 + k_2 + k_3 + k_4 + k_5 + \sqrt{3}  k_6)\\
\frac{1}{2} ( - k_1 + k_2 + k_3 - k_4 + k_5 - \sqrt{3}  k_6) \; ,\quad \frac{1}{2} ( - k_1 + k_2 + k_3 - k_4 - k_5 + \sqrt{3}  k_6) \\
\end{array}
\end{equation}

All these vectors have projection $\frac{1}{3}(k_2 + k_3 - 2 k_1)$ on the plane $\mathbf{\Pi}$.

The other 2 $\jo_3^2$'s are obtained by a cyclic permutation of the indexes of  $k_1$, $k_2$, $k_3$. The 3  $\job_3^2$'s are obtained by reversing the roots' sign.

\subsubsection{The roots associated with $\mathbf{g_{II}} =  \mathbf{g_0^2}$}

The 12 roots of   $\mathbf{g_0^2} = \adad \equiv \adadud$ are 

\[
\begin{array}{l}
\mathbf{a_2 ^{(1)}} \, : \\
\pm (k_4 + k_5) \\ \pm \frac{1}{2} ( k_1 + k_2 + k_3 - k_4 - k_5 - \sqrt{3}  k_6)\; , \quad  \pm\frac{1}{2} ( k_1 + k_2 + k_3 + k_4 + k_5 - \sqrt{3}  k_6)
\end{array}
\]

\[
\begin{array}{l}
\mathbf{a_2 ^{(2)}} \, :\\
\pm (k_4 - k_5) \\ \pm \frac{1}{2} ( k_1 + k_2 + k_3 - k_4 + k_5 + \sqrt{3}  k_6)\; , \quad  \pm \frac{1}{2} ( k_1 + k_2 + k_3 + k_4 - k_5 + \sqrt{3}  k_6)
\end{array}
\]

\subsubsection{The geometry of the Jordan pair $\mathbf V$ and of  $\mathbf{g_{III}}=\mathbf{g_0^2} \oplus \cc \oplus \mathbf V$}

The roots of $\adu$ lie on the plane
\[
\mathbf\Pi^0_{(1)} = \left\{ s\ \frac{\sqrt{2}}{2} (k_4 + k_5) + t\ \frac{\sqrt{6}}{6} (k_1 + k_2 + k_3 - \sqrt{3} k_6) \right\}
\]
and those of $\add$ lie on the orthogonal plane
\[
\mathbf\Pi^0_{(2)} = \left\{ s\ \frac{\sqrt{2}}{2} (k_4 - k_5) + t\ \frac{\sqrt{6}}{6} (k_1 + k_2 + k_3 + \sqrt{3} k_6) \right\}
\]

The root vectors of the h.w. $\jo_3^2$  all lie on three parallel plane defined by the linear spans:
\[
\mathbf\Pi^+_{(1) i} = \left\{\mathbf{u_i} + s\ \frac{\sqrt{2}}{2} (k_4 + k_5) + t\ \frac{\sqrt{6}}{6} (k_1 + k_2 + k_3 - \sqrt{3} k_6) \right\}\; , \quad i = 1,2,3
\]
where the three vectors $\mathbf{u_i}$, orthogonal to $\mathbf\Pi^0_{(1)}$ are:
\begin{align*}
\mathbf{u_1} &= \frac{1}{6} (- 5 k_1 + k_2 + k_3 +3 k_4 - 3 k_5 - \sqrt{3} k_6) \\
\mathbf{u_2} &= \frac{1}{3} (- k_1 + 2 k_2 + 2 k_3 + \sqrt{3} k_6) \\
\mathbf{u_3} &= \frac{1}{6} (- 5 k_1 + k_2 + k_3 -3 k_4 + 3 k_5 - \sqrt{3} k_6) \\
\end{align*}

They also lie on another triple of parallel planes:
\[
\mathbf\Pi^+_{(2) i} = \left\{\mathbf{v_i} + s'\ \frac{\sqrt{2}}{2} (k_4 - k_5) + t'\ \frac{\sqrt{6}}{6} (k_1 + k_2 + k_3 + \sqrt{3} k_6) \right\}\; , \quad i = 1,2,3
\]
where the three vectors $\mathbf{v_i}$, orthogonal to $\mathbf\Pi^0_{(2)}$ are:
\begin{align*}
\mathbf{v_1} &= \frac{1}{6} (- 5 k_1 + k_2 + k_3 +3 k_4 + 3 k_5 + \sqrt{3} k_6) \\
\mathbf{v_2} &= \frac{1}{3} (- k_1 + 2 k_2 + 2 k_3 - \sqrt{3} k_6) \\
\mathbf{v_3} &= \frac{1}{6} (- 5 k_1 + k_2 + k_3 -3 k_4 - 3 k_5 + \sqrt{3} k_6) \\
\end{align*}

In table 3 we list the 9 roots of the h.w. $\jo_3^2$ according to the planes they belong to, with the related $(s,t)$ and $(s',t')$ quantum numbers.

The planes $\mathbf\Pi^-_{(k) i}$, for $k=1,2$ and $i=1,2,3$, of the conjugate Jordan algebra, forming a Jordan pair with the h.w. $\jo_3^2$, are obtained by reversing the signs of the vectors $\mathbf{u_i}$ and $\mathbf{v_i}$, giving rise to opposite quantum numbers.

\begin{table}
\begin{equation*}
\begin{array}{|c|c|c|c|c|}
\hline
\mathbf\Pi^-_{(1) i} & \mathbf\Pi^-_{(2) i}&\text{roots}&(s,t) & (s',t')\\
\hline
 & \mathbf\Pi^-_{(2) 1}&- k_1 + k_4 & \left( \frac{\sqrt{2}}{2},-\frac{\sqrt{6}}{6}\right) &\\
\mathbf\Pi^-_{(1) 1} & \mathbf\Pi^-_{(2) 2}&\frac{1}{2}(- k_1 + k_2 + k_3 + k_4 - k_5 - \sqrt{3} k_6) &\left(0, \frac{\sqrt{6}}{3}\right)& \left( \frac{\sqrt{2}}{2},- \frac{\sqrt{6}}{6}\right) \\
 & \mathbf\Pi^-_{(2) 3}&- k_1 - k_5 & \left( -\frac{\sqrt{2}}{2},-\frac{\sqrt{6}}{6}\right) &\\
\hline
 & \mathbf\Pi^-_{(2) 1}&\frac{1}{2}(- k_1 + k_2 + k_3 + k_4 + k_5 + \sqrt{3} k_6)& \left( \frac{\sqrt{2}}{2},-\frac{\sqrt{6}}{6}\right) & \\
\mathbf\Pi^-_{(1) 2} & \mathbf\Pi^-_{(2) 2}&k_2 + k_3&\left(0, \frac{\sqrt{6}}{3}\right)& \left(0, \frac{\sqrt{6}}{3}\right) \\
 & \mathbf\Pi^-_{(2) 3}&\frac{1}{2}(- k_1 + k_2 + k_3 - k_4 - k_5 + \sqrt{3} k_6)  & \left( -\frac{\sqrt{2}}{2},-\frac{\sqrt{6}}{6}\right) &\\
\hline
 & \mathbf\Pi^-_{(2) 1}&- k_1 + k_5 & \left( \frac{\sqrt{2}}{2},-\frac{\sqrt{6}}{6}\right) &\\
\mathbf\Pi^-_{(1) 3} & \mathbf\Pi^-_{(2) 2}&\frac{1}{2}(- k_1 + k_2 + k_3 + k_4 - k_5 - \sqrt{3} k_6) &\left(0, \frac{\sqrt{6}}{3}\right)& \left( -\frac{\sqrt{2}}{2},- \frac{\sqrt{6}}{6}\right) \\
 & \mathbf\Pi^-_{(2) 3}&- k_1 - k_4 & \left( -\frac{\sqrt{2}}{2},-\frac{\sqrt{6}}{6}\right) &\\
\hline
\end{array}
\end{equation*}
\begin{center} Table 3. $\adad$ quantum numbers of the h.w. $\jo_3^2$.\end{center}
\end{table}

All roots of $\jo_3^2$ lie on the following 4-dimensional space $\mathbf{\Sigma^+}$ parallel to the 4-dimensional space $\mathbf{\Sigma^0}$ of the roots of $\adad$ at a distance $\frac{\sqrt{6}}{3}$:
\begin{align*}
\mathbf{\Sigma^+} = \left\{  \frac{1}{3} (k_2 + k_3 - 2 k_1) + s\ \frac{\sqrt{2}}{2} (k_4 + k_5) + t\ \frac{\sqrt{6}}{6} (k_1 + k_2 + k_3 - \sqrt{3} k_6) \right. \\ \left.+  s'\ \frac{\sqrt{2}}{2} (k_4 - k_5) + t'\ \frac{\sqrt{6}}{6} (k_1 + k_2 + k_3 + \sqrt{3} k_6) \right\}
\end{align*}
 The conjugate $\job_3^2$ lies on the 4-dimensional space parallel to that with the vector $ -\frac{1}{3} (k_2 + k_3 - 2 k_1)$ replacing $\frac{1}{3} (k_2 + k_3 - 2 k_1)$, at the same distance $\frac{\sqrt{6}}{3}$ from $\mathbf{\Sigma^0}$. They form a Jordan pair as can be easily proven by the same argument used in the case of $\fq$. The set of roots corresponding to this Jordan pair plus those of $\adad$ form indeed the root system of $\mathbf{a_5}$, which is three graded like the other Lie algebras in the third row of the magic square.

\noindent\textbf{Roots of $\mathbf{a_5} = \adad \oplus \cc \oplus(\jo_3^2,\job_3^2)$}

Let $\{ \eta_i \, , \; i=1,\dots ,6\}$ be a basis in $\rr^6$ such that the roots of $\mathbf{a_5}$ can be written as $\eta_i - \eta_j\, , \; i \not= j = 1,\dots , 6$.

We make the following correspondence:
\[
\begin{array}{rl}
\adu & \\
\eta_1 - \eta_2 &= k_4 + k_5 \\ 
\eta_3 - \eta_1 &= \frac{1}{2} ( k_1 + k_2 + k_3 - k_4 - k_5 - \sqrt{3}  k_6)\\
\eta_2 - \eta_3 &= \frac{1}{2} (- k_1 - k_2 - k_3 - k_4 - k_5 + \sqrt{3}  k_6) \\
\add & \\
\eta_4 - \eta_5 &= \frac{1}{2} ( k_1 + k_2 + k_3 - k_4 + k_5 + \sqrt{3}  k_6) \\ 
\eta_6 - \eta_4 &=  k_4 - k_5 \\
\eta_5 - \eta_6 &= \frac{1}{2} (-k_1 - k_2 - k_3 - k_4 + k_5 - \sqrt{3}  k_6)
\end{array}\]
\[
\begin{array}{rl}
\jo_3^2 & \\
\eta_1 - \eta_4 &=  - k_1 +  k_4 \\
\eta_1 - \eta_5 &=  \frac{1}{2} ( -k_1 + k_2 + k_3 + k_4 + k_5 + \sqrt{3}  k_6)\\ 
\eta_1 - \eta_6 &=  - k_1 + k_5\\
\eta_2 - \eta_4 &=  - k_1 -  k_5\\
\eta_2 - \eta_5 &=  \frac{1}{2} ( -k_1 + k_2 + k_3 - k_4 - k_5 + \sqrt{3}  k_6)\\ 
\eta_2 - \eta_6 &=  - k_1 - k_4\\
\eta_3 - \eta_4 &=  \frac{1}{2} ( -k_1 + k_2 + k_3 + k_4 - k_5 - \sqrt{3}  k_6)\\ 
\eta_3 - \eta_5 &=   k_2 + k_3\\
\eta_3 - \eta_6 &=  \frac{1}{2} ( -k_1 + k_2 + k_3 - k_4 + k_5 - \sqrt{3}  k_6)\\ 
\end{array}
\]
The roots with opposite sign complete the roots of $\mathbf{a_5}$ with the rest of the roots of $\adad$ and those of $\job_3^2$ 

By a cyclic permutation of  the indexes of  $k_1$, $k_2$, $k_3$ we obtain an analogous result for the other two Jordan pairs, all sharing the same $\adad$ roots for the algebra $\mathbf{g_0^2}$, but with different orientations of the axis defining $\cc$ along the vectors  $\frac{1}{3} (k_1 + k_3 - 2 k_2)$  and  $\frac{1}{3} (k_1 + k_2 - 2 k_3)$ . We get in six dimensions three copies of $\mathbf{a_5}$ all sharing the same $\adad$. All the spaces spanned by the three Jordan pairs are parallel to the space $\mathbf{\Sigma^0}$, and all at the same distance from it. Notice that in six dimensions there is an infinite number of 4-dimensional spaces parallel to a given one, all at the same distance from it. This can be easily checked as we have done for $\fq$.

\subsubsection{Nested Jordan pairs}

If we dig inside $\mathbf{g_0^2}$ we find another Jordan Pair plus the Lie algebra of its automorphism group: these are a  $(2, \bar 2)$ of $\mathbf{a_1}$ plus two $\mathbf{a_1} \oplus \cc$ making up, all together,  $\adad$.

\subsection{\Large $\;\;\est$}

\subsubsection{The roots associated with the h.w. Jordan algebra $\jo_3^4$}

The 15 roots corresponding to the highest weight $\jo_3^4$ in Fig. 3 are
\begin{equation}
\begin{array}{l}
- k_1 \pm k_4  \; , \qquad - k_1 \pm k_5 \; , \qquad - k_1 \pm k_6 \; , \qquad k_2 + k_3\\
\frac{1}{2} ( - k_1 + k_2 + k_3 - k_4 - k_5 - k_6 \pm \sqrt{2}  k_7)  \; ,\quad \frac{1}{2} ( - k_1 + k_2 + k_3 - k_4 + k_5 + k_6 \pm \sqrt{2}  k_7\\
\frac{1}{2} ( - k_1 + k_2 + k_3 + k_4 - k_5 - k_6 \pm \sqrt{2}  k_7)  \; ,\quad \frac{1}{2} ( - k_1 + k_2 + k_3 + k_4 + k_5 - k_6 \pm \sqrt{2}  k_7)\\
\end{array}
\label{j34}\end{equation}

all these vectors have projection $\frac{1}{3}(k_2 + k_3 - 2 k_1)$ on the plane $\mathbf\Pi$ in Fig 3.

The 15 roots of the conjugate  $\job_3^4$, forming a Jordan pair with the h.w. $\jo_3^4$, are obtained by reversing the signs.

The other two Jordan pairs are obtained by a cyclic permutation of the indexes of  $k_1$, $k_2$, $k_3$.

\subsubsection{The roots associated with $\mathbf{g_{II}} =  \mathbf{g_0^4}$}

The 30 roots of   $\mathbf{g_0^4} = \mathbf{a_5}$ are 
\[
\begin{array}{l}
\pm k_4 \pm k_5  \; , \quad \pm k_4 \pm k_6 \; , \quad \pm k_5 \pm k_6 \; , \quad \pm \sqrt{2}  k_7\\ 
\pm \frac{1}{2} ( k_1 + k_2 + k_3 \pm k_4 \pm k_5 \pm k_6 \pm \sqrt{2}  k_7) \quad \text{even number of} \; +\frac{1}{2} \\
\end{array}
\]

We now check that these 30 roots describe indeed $\mathbf{a_5}$ and that, together with the 15 roots \eqref{j34} plus the 15 of opposite sign, they form the root diagram of $\mathbf{d_6}$.

\vspace{.4cm}
\noindent\textbf{Roots of $\mathbf{d_6} = \mathbf{a_5} \oplus \cc \oplus (\jo_3^4,\job_3^4)$}
\vspace{.4cm}

Let $\{ \eta_i \, , \; i=1,\dots ,6\}$ be the following orthonormal basis of $\rr^6$. The 60 roots of $\mathbf{d_6}$ can be written as:
\[ \pm \eta_i \pm \eta_j \; , \quad i \not = j = 1, \dots , 6\]

In order to write the roots of  the $\mathbf{d_6}$ sub-algebra inside $\est$ we define:
\[
\begin{array}{rl}
\eta_1 &= \frac{1}{2} ( - k_1 -  k_4 - k_5 -  k_6)\\
\eta_2 &= \frac{1}{2} ( - k_1 -  k_4 + k_5 +  k_6)\\
\eta_3 &= \frac{1}{2} ( - k_1 +  k_4 - k_5 +  k_6)\\
\eta_4 &= \frac{1}{2} ( - k_1 +  k_4 + k_5 -  k_6)\\
\eta_5 &= \frac{1}{2} (k_2 + k_3 + \sqrt{2}  k_7) \\
\eta_6 &= \frac{1}{2} (k_2 + k_3 - \sqrt{2}  k_7) 
\end{array}
\]
Then the roots of $\mathbf{a_5}$  can be written as $\eta_i - \eta_j\, , \; i \not= j = 1,\dots , 6$:
\[
\begin{array}{rl}
\eta_1 - \eta_2 &= - k_5 - k_6 \\ 
\eta_1 - \eta_3 &= - k_4 - k_6 \\
\eta_1 - \eta_4 &= - k_4 - k_5 \\
\eta_1 - \eta_5 &= \frac{1}{2} (- k_1 - k_2 - k_3 - k_4 - k_5 - k_6 - \sqrt{2}  k_7) \\
\eta_1 - \eta_6 &= \frac{1}{2} (- k_1 - k_2 - k_3 - k_4 - k_5 - k_6 + \sqrt{2}  k_7) \\
\eta_2 - \eta_3 &= - k_4 + k_5 \\ 
\eta_2 - \eta_4 &= - k_4 + k_6 \\
\eta_2 - \eta_5 &= \frac{1}{2} (- k_1 - k_2 - k_3 - k_4 + k_5 + k_6 - \sqrt{2}  k_7) \\
\eta_2 - \eta_6 &= \frac{1}{2} (- k_1 - k_2 - k_3 - k_4 + k_5 + k_6 + \sqrt{2}  k_7) \\
\eta_3 - \eta_4 &=  - k_5 + k_6 \\
\eta_3 - \eta_5 &= \frac{1}{2} (- k_1 - k_2 - k_3 + k_4 - k_5 + k_6 - \sqrt{2}  k_7) \\
\eta_3 - \eta_6 &= \frac{1}{2} (- k_1 - k_2 - k_3 + k_4 - k_5 + k_6 + \sqrt{2}  k_7) \\
\eta_4 - \eta_5 &= \frac{1}{2} (- k_1 - k_2 - k_3 + k_4 + k_5 - k_6 - \sqrt{2}  k_7) \\
\eta_4 - \eta_6 &= \frac{1}{2} (- k_1 - k_2 - k_3 + k_4 + k_5 - k_6 + \sqrt{2}  k_7) \\
\eta_5 - \eta_6 &=  \sqrt{2}  k_7 \\
\end{array}
\]
The other 15 roots of $\mathbf{a_5}$  are those of opposite sign.

The roots associated with the h.w. $\jo_3^4$ can be written as $\eta_i + \eta_j\, , \; i \not= j = 1,\dots , 6$:
\[
\begin{array}{rl}
\eta_1 + \eta_2 &= - k_1 - k_4 \\ 
\eta_1 + \eta_3 &= - k_1 - k_5 \\
\eta_1 + \eta_4 &= - k_1 - k_6 \\
\eta_1 + \eta_5 &= \frac{1}{2} (- k_1 + k_2 + k_3 - k_4 - k_5 - k_6 + \sqrt{2}  k_7) \\
\eta_1 + \eta_6 &= \frac{1}{2} (- k_1 + k_2 + k_3 - k_4 - k_5 - k_6 - \sqrt{2}  k_7) \\
\eta_2 + \eta_3 &= - k_1 + k_6 \\ 
\eta_2 + \eta_4  &= - k_1 + k_5 \\
\eta_2 + \eta_5 &= \frac{1}{2} (- k_1 + k_2 + k_3 - k_4 + k_5 + k_6 + \sqrt{2}  k_7) \\
\eta_2 + \eta_6 &= \frac{1}{2} (- k_1 + k_2 + k_3 - k_4 + k_5 + k_6 - \sqrt{2}  k_7) \\
\eta_3 + \eta_4 &=  - k_1 + k_4 \\
\eta_3 + \eta_5 &= \frac{1}{2} (- k_1 + k_2 + k_3 + k_4 - k_5 + k_6 + \sqrt{2}  k_7) \\
\eta_3 + \eta_6 &= \frac{1}{2} (- k_1 + k_2 + k_3 + k_4 - k_5 + k_6 - \sqrt{2}  k_7) \\
\eta_4 + \eta_5 &= \frac{1}{2} (- k_1 + k_2 + k_3 + k_4 + k_5 - k_6 + \sqrt{2}  k_7) \\
\eta_4 + \eta_6 &= \frac{1}{2} (- k_1 + k_2 + k_3 + k_4 + k_5 - k_6 - \sqrt{2}  k_7) \\
\eta_5 + \eta_6 &=  k_2 + k_3 \\
\end{array}
\]

The roots with opposite sign, namely the rest of the roots of $\mathbf{a_5}$ and those of $\job_3^2$,  complete the roots of $\mathbf{d_6}$.

\subsubsection{The geometry of the Jordan pair $\mathbf V$ and of  $\mathbf{g_{III}}=\mathbf{g_0^4} \oplus \cc \oplus \mathbf V$}

The roots of $\mathbf{a_5}$  lie in the 5-dimensional space that we can write as the linear span of orthogonal vectors:

\[
\mathbf{\Sigma^0} = \{ s_1\ k_4 +  s_2\ k_5 + s_3\ k_6 + s_4\ k_7 + s_5\ (k_1 + k_2 + k_3) \}
\]

The roots \eqref{j34} of the h.w. $\job_3^4$ lie in the in a parallel 5 dimensional space:

\[
\mathbf{\Sigma^+} = \{\frac{1}{3}(k_2 + k_3 - 2 k_1) +  s_1\ k_4 +  s_2\ k_5 + s_3\ k_6 + s_4\ k_7 + s_5\ (k_1 + k_2 + k_3) \}
\]

The Jordan pair is $(\jo_3^4, \job_3^4)$. The Lie Algebra of its automorphism group is $\mathbf{a_5} \oplus \cc$ where $\cc$ is the axis along the vector $\frac{1}{3}(k_2 + k_3 - 2 k_1)$. As already remarked, all root vectors of $\jo_3^4$ (resp. $\job_3^4$) project on the point $\frac{1}{3}(k_2 + k_3 - 2 k_1)$ (resp. $- \frac{1}{3}(k_2 + k_3 - 2 k_1)$) in the plane of Fig. 3, hence they all carry the same $\cc$  quantum number $\pm\frac{\sqrt{6}}{3}$.

By a cyclic permutation of  the indexes of  $k_1$, $k_2$, $k_3$ we obtain an analogous result for the other two Jordan pairs, all sharing the same $\mathbf{a_5}$ roots for the algebra $\mathbf{g_0^4}$, but with different orientations of the axis defining $\cc$ along the vectors  $\frac{1}{3} (k_1 + k_3 - 2 k_2)$  and  $\frac{1}{3} (k_1 + k_2 - 2 k_3)$ . We get in seven dimensions three copies of $\mathbf{d_6}$ all sharing the same $\mathbf{a_5}$. All the spaces spanned by the three Jordan pairs are parallel to the space $\mathbf{\Sigma^0}$, and all at the same distance from it. Notice that in seven dimensions there is an infinite number of 5-dimensional spaces parallel to a given one, all at the same distance from it. This can be easily checked as we have done for $\fq$.

\subsubsection{Nested Jordan pairs}

If we dig inside $\mathbf{g_0^4} = \mathbf{a_5}$ we find the Jordan Pair  $(\jo_3^2, \job_3^2) = (3\times 3 , \bar 3 \times \bar 3)$ plus the Lie algebra of its automorphism group $\adad \oplus \cc$ described in the previous case of $\es$.

\subsection{\Large $\;\;\eo$}

\subsubsection{The roots associated with the h.w. Jordan algebra $\jo_3^8$}

The 27 roots corresponding to the highest weight $\jo_3^8$ in Fig. 3 are:

\begin{equation}
\begin{array}{l}
 -k_1 \pm  k_j \; , \quad  j = 4, \dots ,8   \quad ;\quad k_2+k_3\\
\frac{1}{2} (- k_1 + k_2 + k_3 \pm k_4  \pm k_5  \pm k_6 \pm k_7 \pm k_8)\qquad \text{even number of + signs}
\end{array}
\label{j38}\end{equation}
all these vectors have projection $\frac{1}{3}(k_2 + k_3 - 2 k_1)$ on the plane $\mathbf\Pi$ in Fig 3.

The 27 roots of the conjugate  $\job_3^8$, forming a Jordan pair with the h.w. $\jo_3^8$, are obtained by reversing the signs.

The other two Jordan pairs are obtained by a cyclic permutation of the indexes of  $k_1$, $k_2$, $k_3$.

\subsubsection{The roots associated with $\mathbf{g_{II}} =  \mathbf{g_0^8}$}

The 72 roots of   $\mathbf{g_0^8} = \es$ are 
\begin{equation}
\begin{array}{l}
\pm k_i \pm  k_j \; , \quad  i,j = 4, \dots ,8 \\
\pm \frac{1}{2} (k_1 + k_2 + k_3 \pm k_4  \pm k_5  \pm k_6 \pm k_7 \pm k_8)\qquad \text{even number of + signs}
\end{array}
\end{equation}

We can easily recognize the roots of $\es$ in Table 2 by introducing $k'_i = k_i+3$, $i=1,\dots,5$ and $k'_6 = -\frac{\sqrt{3}}{3}(k_1+k_2+k_3)$. The roots of  $\est$ can be recognized by setting $\tilde k_i = k_i+3$, $i=1,\dots,5$, $\tilde k_6 = k_1$ and  $\tilde k_7 = \frac{\sqrt{2}}{2}(k_2+k_3)$.

\subsubsection{The geometry of the Jordan pair $\mathbf V$ and of  $\mathbf{g_{III}}=\mathbf{g_0^8} \oplus \cc \oplus \mathbf V$}

The roots of $\es$  lie in the 6-dimensional space that we can write as the linear span of orthogonal vectors:

\[
\mathbf{\Sigma^0} = \{ s_1\ k_4 +  s_2\ k_5 + s_3\ k_6 + s_4\ k_7 + s_5\ k_8 + s_6\ (k_1 + k_2 + k_3) \}
\]

The roots \eqref{j38} of the h.w. $\job_3^8$ lie in the in a parallel 6-dimensional space:

\[
\mathbf{\Sigma^+} = \{\frac{1}{3}(k_2 + k_3 - 2 k_1) +  s_1\ k_4 +  s_2\ k_5 + s_3\ k_6 + s_4\ k_7 + s_5\ k_8 + s_6\ (k_1 + k_2 + k_3)  \}
\]

The Jordan pair is $(\jo_3^8, \job_3^8)$. The Lie Algebra of its automorphism group is $\es \oplus \cc$ where $\cc$ is the axis along the vector $\frac{1}{3}(k_2 + k_3 - 2 k_1)$. As already remarked, all root vectors of $\jo_3^8$ (resp. $\job_3^8$) project on the point $\frac{1}{3}(k_2 + k_3 - 2 k_1)$ (resp. $- \frac{1}{3}(k_2 + k_3 - 2 k_1)$) in the plane of Fig. 3, hence they all carry the same $\cc$  quantum number $\pm\frac{\sqrt{6}}{3}$.

By a cyclic permutation of  the indexes of  $k_1$, $k_2$, $k_3$ we obtain an analogous result for the other two Jordan pairs, all sharing the same $\es$ roots for the algebra $\mathbf{g_0^8}$, but with different orientations of the axis defining $\cc$ along the vectors  $\frac{1}{3} (k_1 + k_3 - 2 k_2)$  and  $\frac{1}{3} (k_1 + k_2 - 2 k_3)$ . We get in eight dimensions three copies of $\est$ all sharing the same $\es$. All the spaces spanned by the three Jordan pairs are parallel to the space $\mathbf{\Sigma^0}$, and all at the same distance from it. Notice that in eight dimensions there is an infinite number of 6-dimensional spaces parallel to a given one, all at the same distance from it. This can be easily checked as we have done for $\fq$.

\subsubsection{Nested Jordan pairs}

If we dig inside $\mathbf{g_0^8} = \es$ we find three Jordan Pairs, each of the type  $(\jo_3^2, \job_3^2) = (3\times 3 , \bar 3 \times \bar 3)$, plus the Lie algebra of the automorphism group of each of them $\adad \oplus \cc$ described in the previous case of $\es$.

We thus identify four different  $\ad$'s within $\eo$ plus 6 Jordan pairs. Giving different superscripts to the four $\ad$'s we have:

\begin{equation}
\begin{array}{rl}
\eo &= \adc \oplus 3\times (\jo_3^8 , \job _3^8 ) \oplus \mathbf{g_0^8} \\
      &= \adc \oplus 3\times (\jo_3^8 , \job _3^8 ) \oplus \adf \oplus 3\times (\jo_3^2 , \job _3^2 ) \oplus \mathbf{g_0^2} \\
      &= \adc \oplus 3\times (\jo_3^8 , \job _3^8 ) \oplus \adf \oplus 3\times (\jo_3^2 , \job _3^2 ) \oplus \adgu \oplus \adgd
\end{array}
\label{deo}\end{equation}

\section{Representations}

I briefly sketch in this section a possible representation of the $\eo$  algebra which exhibits its Jordan pair content.

The way I would represent $\eo$ is a development of the representation of $\est$ through generalized $2\times 2$ matrices, shown in \cite{tob}. The starting point of that paper is the representation of the quaternion algebra through the Pauli matrices, which leads directly to the three grading of $\est$. In the case of $\eo$ a suitable representation of the octonions is via the Zorn matrices, \cite{zorn,loos2} which exhibit the $(3, \bar 3)$ structure that we can extend to the Jordan pair content of $\eo$ and to the action on the $(3, \bar 3)$ modules of the \emph{external} $\ad$ in Figure 3.

The guideline goes as follows:

\begin{itemize}
\item Represent the octonions as Zorn matrices
\item Extend the Zorn matrices to represent $Der(\oo) = \gd$
\item Combine the extended Zorn matrices with the Tits construction \eqref{titsms}
\item Decompose the representation of $\es$ to finally get $\eo$ in terms of Jordan pairs and $\ad$'s only
\end{itemize}

If $a \in \oo$ we write $a = a_0 + \sum_{k=1}^7{a_k u_k}$ where $a_\ell \in \cc$ for $\ell = 0, \dots, 7$ and $u_1,\dots ,u_7$ are the octonionic imaginary units.
 
Let us denote by $i$ the imaginary unit in $\cc$. We introduce 2 idempotent elements
\[ \rho_\pm = \frac{1}{2} (1 \pm i u_7)\]
and 6 nilpotent elements
\[\ekpm = \rho^\pm u_k \, , \; k=1,2,3\]

The Zorn representation of $a\in \oo$ is:
\begin{equation}
a = \alpha^+ \rho^+ + \alpha^- \rho^- + \sum_k (\alpha_k^+ \ekp + \alpha_k^- \ekm) \leftrightarrow \left[ \begin{matrix}\alpha^+ & A^+ \\ A^- & \alpha^-\end{matrix}\right]
\end{equation}
where $A^\pm \in \cc^3$ have vector components $\alpha_k^\pm\, , \; k=1,2,3$ and the octonionic multiplication is a generalization of the matrix multiplication:
\begin{equation}
\begin{array}{ll}
a b & \leftrightarrow \left[ \begin{matrix}\alpha^+ & A^+ \\ A^-  & \alpha^-\end{matrix}\right]  \left[ \begin{matrix}\beta^+ & B^+ \\ B^- & \beta^-\end{matrix}\right] \\ 
\\
&= \left[ \begin{matrix}\alpha^+ \beta^+ + A^+ \cdot B^- & \alpha^+ B^+ + \beta^- A^+ + A^- \times B^-\\ \alpha^- B^- + \beta^+ A^- + A^+ \times B^+  & \alpha^- \beta^- + A^-\cdot B^+\end{matrix}\right]
\end{array}
\end{equation}
with $A^\pm \cdot B^\mp = - \alpha_k^\pm \beta_k^\mp$ and $A,B \to A \times B$ is the standard vector product in $\cc^3$.

The next step is to write the Lie algebra $\gd$ using an extension of the Zorn matrices and their multiplication rule with an $\ad$ matrix replacing $\alpha^+$. This representation shows $\gd$ as $\ad$ plus its modules $(3, \bar 3)$.

Finally, let me outline how the Tits construction fits into this picture. The idea is to write:
\begin{equation}
\eo = Der(\oo) \oplus \oo_0 \otimes \jo_0^8 \oplus Der( \jo^8) = \lk_0 \oplus \sum_{\pm k}{\lk_{\pm k}} \, ,\; k=1,2,3
\end{equation}
where
\[ \lk_0 =  D_7 \oplus i u_7 \otimes \jo_0^8 \oplus Der(\jo^8) \; \text{and} \; \lk_{\pm k} = d_k^\pm D_k ^\pm \oplus \alpha_k^\pm \ekpm \otimes \jo_0^8 \, , \; d_k^\pm , \, \alpha_k^\pm \in \cc
\]
Here $\jo^8 \equiv \jo_3^8$, $\jo_0^8$ is a traceless $\jo_3^8$ matrix; $D_7 = \ad$ is the subalgebra of derivations leaving the imaginary unit $u_7$ fixed;  $D_k^\pm = \pm \frac{3}{2} D_{i u_7,\ekpm}$ is a derivation:
\[ D_{a,b}\  c = \frac{1}{3}[[a,b],c] - (a,b,c) \; ; \quad (a,b,c)=(ab)c - a(bc)\]

We identify $a\otimes x$ with $a_z\otimes x$, where $a_z$ is the Zorn matrix representation of $a$ and $Der_k^\pm$ with the corresponding Zorn matrix representation of $\ekpm$ . We use the complex parameters $d_k^\pm$ in order to provide the trace to $\jo_0$.

The sequence of implications, starting from the Tits construction would be like this:
\begin{equation}
\begin{array}{ll}
\eo &= Der(\oo) \oplus \oo_0 \otimes \jo_0^8 \oplus Der( \jo^8) \\ &=  \adc \oplus \alpha_k^\pm \ekpm \otimes \jo_0^8 \oplus d_k^\pm  Der_k^\pm \oplus (i u_7) \otimes \jo_0^8 \oplus Der( \jo^8) \\
& = \adc \oplus \alpha_k^\pm \ekpm \otimes \jo^8 \oplus Der(\oo) \oplus\oo_0 \otimes \jo_0^2 \oplus Der(\jo^2) \\
&=\adc \oplus \alpha_k^\pm \ekpm \otimes \jo ^8 \oplus \adf \oplus \alpha_k^\pm \ekpm \otimes \jo^2  \\ &\phantom{=}\;\, \oplus (i u_7) \otimes \jo_0^2 \oplus Der( \jo^2) \\
&= \adc \oplus \adf \oplus \adgu \oplus \adgd \oplus 3 \times ({\jo^8 ,\job^8) \oplus 3 \times (\jo^2 ,\job^2)}
\end{array}
\end{equation}

Work is still in progress along these lines and will appear in a forthcoming paper. 

\section{Elementary particle physics}

If we look at the decomposition \eqref{deo} we are led to interpret the labels  \textbf{c} as color and \textbf{f} as flavor. 
	In this interpretation the three pairs $(\jo_3^8 , \job _3^8 )$ accommodate the quarks in three colors of particles-antiparticles, whereas the three pairs $(\jo_3^2 , \job _3^2 )$ sitting in the colorless $\mathbf{g_0^8}$ accommodate the three families of leptons-antileptons. 
	Including spin, each particle must appear with four different degrees of freedom: left (up and down) and right (up and down), except, possibly, for the neutrino which could be a Majorana neutrino and be only left-handed.  
	We can therefore put six (quarks, antiquarks) in a  (say) blue $(\jo_3^8 , \job _3^8 )$ . We can make them coincide with three octonions: one for blue up-down quarks, one for blue charm-strange quarks, one for blue top-bottom quarks. We are left with 3 extra degrees of freedom.
	In the same fashion, we can put a family of leptons-antileptons pairs in $(\jo_3^2 , \job _3^2 )$ by letting the six off-diagonal degrees of freedom of be the electron and a Majorana neutrino, and analogously for the families of the muon and  leptons. Again we are left with 3 extra degrees of freedom, which reduce to only 1 in case right handed neutrinos are included.

Let us review the explicit form of the roots according to this interpretation. 

\vspace{.3cm}
\noindent\textbf{quarks of color c = 1,2,3} (corresponding antiquarks have reversed signs)
\begin{equation*}
\begin{array}{l}
 -k_c \pm  k_j \; , \quad  j = 4, \dots ,8   \quad ;\quad -k_c + k_1 + k_2+k_3\\
- k_c + \frac{1}{2} ( k_1 + k_2 + k_3 \pm k_4  \pm k_5  \pm k_6 \pm k_7 \pm k_8)\qquad \text{even number of + signs}
\end{array}
\end{equation*}

\vspace{.3cm}
\noindent\textbf{leptons in the family f = 4,5,6} (corresponding antileptons have reversed signs)
\begin{equation*}
\begin{array}{l}
 -k_f \pm  k_j \; , \quad  j = 7,8   \quad ;\quad -k_f + k_4 + k_5+k_6\\
- k_f + \frac{1}{2} [ \pm ( k_1 + k_2 + k_3) + k_4  + k_5  + k_6 \pm k_7 \pm k_8]\; \text{even number of + signs}
\end{array}
\end{equation*}

\vspace{.3cm}
\noindent $\adc$ : $\pm (k_i - k_j)$ , $i<j = 1,2,3$

\vspace{.3cm}
\noindent $\adf$ : $\pm (k_i - k_j)$ , $i<j = 4,5,6$

\vspace{.3cm}
\noindent $\adgu$ : $\pm (k_7 + k_8)$, $\pm \frac{1}{2} ( k_1 + k_2 + k_3 + k_4  + k_5  + k_6 - k_7 - k_8)$ \\
\phantom{$\adgu$ : }$\pm \frac{1}{2} ( k_1 + k_2 + k_3 + k_4  + k_5  + k_6 + k_7 + k_8)$

\vspace{.3cm}
\noindent $\adgd$ : $\pm (k_7 - k_8)$ , $\pm \frac{1}{2} ( - k_1 - k_2 - k_3 + k_4  + k_5  + k_6 - k_7 + k_8)$\\
\phantom{$\adgd$ : }$\pm \frac{1}{2} ( - k_1 - k_2 - k_3 + k_4  + k_5  + k_6 + k_7 - k_8)$

\vspace{.3cm}
What physics should a theory with an $\eo$ symmetry describe? Certainly a very high energy physics, far beyond our present experience and our experimental reach. It could relate to a string theory, like the heterotic  one, since we are dealing with a complex Lie algebra hence an $\eo \times \eo$ algebra over $\rr$. It could extend to supersymmetry, although the $\eo$ symmetry is so beautiful as it stands that one should force such an extension into the theory: $\eo$, in the view I am presenting here, shows particle-antiparticle pairs, the Jordan pairs, in the right number of colors and families, plus their symmetries, which in turn are generated by the pairs themselves, through the trilinear map $z^\sigma \to  V_{x^\sigma , y ^{-\sigma}} z^\sigma$. Besides, another peculiarity adds up to the beauty of $\eo$: its lowest dimensional irreducible representation is the adjoint representation.

\begin{figure}
\begin{center}
\includegraphics{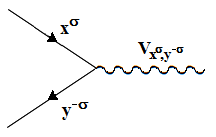}\\ \nopagebreak
\textbf{Fig.6} An elementary interaction, viewed as a Feynman diagram.
\end{center}
\end{figure}

My personal point of view is that, at such a high energy, at or beyond the Planck scale, the picture of spacetime has to be radically changed. I can hardly make any sense of the fact that such an energetic particle is  sitting on a background spacetime, if I think that general relativity taught us that spacetime is in fact dynamical. I would rather view that particle feel only (quantum) interactions, including one that leads to  gravity, to be accommodated within $\adgu\oplus \adgd$. I would still view an \emph{elementary} interaction being described by an elementary Feynman diagram involving the trilinear map, as depicted in Fig. 6, but with no question of point or extended particle, simply because the underlying spacetime geometry is not there: there is only a, let us say, background independent \emph{spectral theory}.

In this view the classical spacetime is a byproduct of the interactions, obtained by taking very rough approximations. It is as far from the interactions exchanged by elementary particles at the Planck scale, as the Planck scale is far from our experience.

The aim of developing along these lines a physical theory that could not possibly relay on any direct confirmation, is to find a consistent quantum theory of gravity together with the other known basic interactions. As Carlo Rovelli says in his book, \cite{carlo}: \emph{the difficulty is not to discriminate among many complete and consistent quantum theories of gravity. We would be content with one}. 

This is, of course, far beyond the scope of the present paper, since no physics has been spoken here besides these mere speculations.
 
\vspace{.7cm}
\noindent\textbf{Acknowledgments} I am very grateful to Prof. V.S.Varadarajan for the scientific advice and the financial support. The discussions we have had together have actually given life to the idea of writing this paper. I also thank prof. Varadarajan and the UCLA Department of Mathematics  for their kind hospitality. The visit to the University of California at Los Angeles was made possible by the \emph{Istituto Nazionale di Fisica Nucleare},  on the grant In. Spec. GE 41.


\end{document}